\documentclass[11pt,twoside]{book}
\usepackage{vshalo}
\usepackage{longtable}
\usepackage{amsmath,amssymb}
\usepackage{graphicx}
\usepackage{lscape}
\usepackage{psfig}
\usepackage{epsf}
\usepackage{index}
\usepackage{natbib}
\usepackage{bigdelim}
\usepackage{multirow}
\makeindex
\newindex{aut}{adx}{and}{\Large Author  Index}
\newcommand{\axindex}[1]{\index[aut]{#1}}

\def\lea{\mathrel{<\kern-1.0em\lower0.9ex\hbox{$\sim$}}}
\def\gea{\mathrel{>\kern-1.0em\lower0.9ex\hbox{$\sim$}}}

\begin{document}


\pagestyle{myheadings}
\setcounter{equation}{0}\setcounter{figure}{0}\setcounter{footnote}{0}\setcounter{section}{0}\setcounter{table}{0} \setcounter{page}{113}

\markboth{Sarajedini \& Yang}{RR Lyrae Variables in the Halo of M\,33}
\title{RR Lyrae Variables in the Halo of M\,33}
\author{Ata Sarajedini and S. -C. Yang}
\axindex{Sarajedini, A.}
\axindex{Yang, S. -C.}
\affil{Department of Astronomy, University of Florida, 211 Bryant Space
Science Center, Gainesville, FL 32611 USA}


\begin{abstract}
The properties of RR Lyrae variables make them excellent probes of
the formation and evolution of a stellar population. The mere presence
of such stars necessitates an age greater than $\sim$10 Gyr while
their periods and amplitudes can be used to estimate the metal
abundance of the cluster or galaxy in which they reside.  These
and other features of RR Lyraes have been used to study the
Local Group late-type spiral galaxy M33. Though these studies
are generally in their infancy, we have established that M33 does
indeed harbor RR Lyraes in its halo and probably also in its disk
suggesting that these two components formed early in the history
of M33. The mean metallicity of the halo RR Lyraes is consistent
with that of the halo globular clusters in M33 at [Fe/H]$\sim$--1.3.
Little is known about the spatial distribution of the RR Lyraes;
this will require wide-field time-series studies with sufficient
photometric depth to allow both the identification of RR Lyraes and
robust period determination.
\end{abstract}

\section{Introduction}
The class of pulsating stars known as RR Lyrae variables are located
at the intersection of the instability strip and the horizontal branch in the
Hertzsprung Russell Diagram. Their utility has been widely documented
in the literature. As such, they can be referred to as the ``swiss army knife" 
of astronomy. 

Because of their low masses ($\sim$0.7 M$_\odot$, Smith 1995), 
the mere presence of RR Lyrae stars in a stellar population 
suggests an old age ($\gea$10 Gyr) 
for the system. As such, one does not need to obtain
deep photometry beyond the old main sequence turnoff in order
to establish the presence of an old population. Generally
speaking, {\it identifying} RR Lyrae stars does not require a
substantial investment of telescope time. 

The periods of the ab-type RR Lyrae stars are related to their
metallicities. These variables pulsate in the fundamental mode
and display a distinctive sawtooth appearance in their light curves.
Using data on field RR Lyraes from
Layden (2005, private communication), Sarajedini et al. (2006)
found 
\begin{eqnarray}
[Fe/H] = -3.43 - 7.82~Log~P_{ab}.
\end{eqnarray}
\noindent The dispersion
in this relation (rms = 0.45 dex) is significant making the determination of 
individual stellar metallicities unreliable, but the relation is useful for 
estimating the mean abundance of a population of RR Lyraes. There 
is a more precise relation given by Alcock et al. (2000) that requires
knowledge of the periods {\it and} amplitudes; with this relation,
the error per star
is reduced to $\sim$0.31 dex and the precision of the resulting abundance
distribution is narrower (Sarajedini et al. 2009). 

Once the metallicities of the RR Lyraes are determined, their
absolute magnitudes can be calculated. The published equations typically
take the form of a linear relation between [Fe/H] and M$_V$(RR). 
A number of different slopes and zeropoints have been derived 
for this equation, but there seems to be convergence on slope
values of $\sim$0.20 and zeropoints of $\sim$0.90 
(Chaboyer 1999, Gratton et al. 2003, 2004; Dotter et al. 2009).

The minimum light colors of ab-type RR Lyraes are largely 
independent of their other properties as shown by Guldenschuh et al.
(2005) and Kunder et al. (2009). 
This is based on a concept originally developed by Sturch (1966).
As a result, if the minimum light colors are well-determined, they can
be compared with $(V-I)_{o,min} = 0.58 \pm 0.02$ and
$(V-R)_{o,min} = 0.28 \pm 0.02$\footnote{Depending on how the
minimum light color is measured, this value could also be 
$(V-I)_{o,min} = 0.27 \pm 0.02$. See Kunder et al. (2009) for 
details.} in order to 
measure the line-of-sight reddening for each star.

As described above, RR Lyrae variables are powerful probes of
the systems in which they reside - star clusters or among the field
populations of galaxies. It is for this reason that studying them in
the Local Group late-type spiral galaxy M33 provides valuable
insights into the properties of this galaxy. Ultimately, we would like to
know how `dwarf spirals' like M33 fit into the process of galaxy formation 
in a Universe dominated by cold dark matter (CDM) with a 
cosmological constant $\Lambda$ (Navarro, Frenk,  \& White 1997).
Comprehensive knowledge of M33's most ancient stars will shed
light on this question. In the remainder of this contribution, we will
describe how RR Lyrae stars have been used to this end.

\section{Previous Studies}

The history of RR Lyrae studies in M33 is relatively short. The earliest
study is that of Pritchet (1988), who presented preliminary results for
a handful of such stars. No data or light curves were shown, but
Pritchet (1988) did estimate a distance of 
(m--M)$_{0}$ = 24.45 $\pm$ 0.2 for M33 based on the RR Lyrae
variables. This value is somewhat smaller than the average of several
different determinations from Galleti et al. (2004) of 
(m--M)$_{0}$ = 24.69 $\pm$ 0.11. 

The first study to unequivocally identify and characterize RR Lyraes in
M33 was that of Sarajedini et al. (2006). They used time-series
observations of two fields in M33 taken with the Wide Field Channel of
the Advanced Camera for Surveys (ACS/WFC) onboard the Hubble
Space Telescope (HST). The observations consisted of 8 epochs 
in the F606W ($\sim$V) filter and 16 epochs in the F814W ($\sim$I)
filter. The data were analyzed using the template-fitting software 
developed by Andrew Layden and described in Layden \& Sarajedini
(2000, see also Mancone \& Sarajedini 2008). Based on these data, 
64 ab-type RR Lyraes were identified.
However, very few c-type variables were uncovered because of their
generally lower amplitude.

The period distribution of the ab-type variables showed two peaks -
one at longer periods which resembles the metal-poor 
RR Lyraes in M3 and M31 (Brown et al. 2004) and one at shorter 
periods which could be from metal-rich RR Lyraes in M33's disk. 
The presence of the latter population is somewhat uncertain given
the recent work of Pritzl et al. (private communication). Sarajedini
et al. (2006) found the mean
metallicity of the metal-poor RR Lyraes to be consistent
with that of halo globular clusters in M33 which have
$\langle$[Fe/H]$\rangle$ = --1.27 $\pm$ 0.11 (Sarajedini et al. 2000).

Figure 12 of Sarajedini et al. (2006) shows that the M33 RR Lyraes are
in their expected location in the color-magnitude diagram (CMD);
in addition, their colors exhibit a dispersion that is consistent
with being significantly affected by differential reddening. This suggests
that some of the RR Lyraes are on the near side of M33 while
others are in the disk or on the far side of the galaxy. One way
to further investigate this possibility is to examine the distribution
of RR Lyrae reddenings using the intrinsic minimum light color
of the ab-types as mentioned in Sec. 1. This analysis was performed
by Sarajedini et al. (2006) and shows that the RR Lyraes span
the range from E(V--I)$<$0.1 up to E(V--I)$\sim$0.7. Given that the
line-of-sight reddening to M33 is E(V--I)$\sim$0.1, this suggests
that RR Lyraes exist in the disk of M33 and in its halo (on the near
and far side). We return to this point later as we discuss the most
recent results on RR Lyraes in M33.

The primary result from the work of Sarajedini et al. (2006) was
that M33 does indeed contain RR Lyrae variables in its halo. This
suggests that the halo of M33 contains some fraction of stars with
ages older than $\sim$10 Gyr. In this regard, the halos of M33,
M31, and the Milky Way are similar. It seems that they started forming
stars at about the same time. 

\begin{figure}
\centerline{\hbox{\psfig{figure=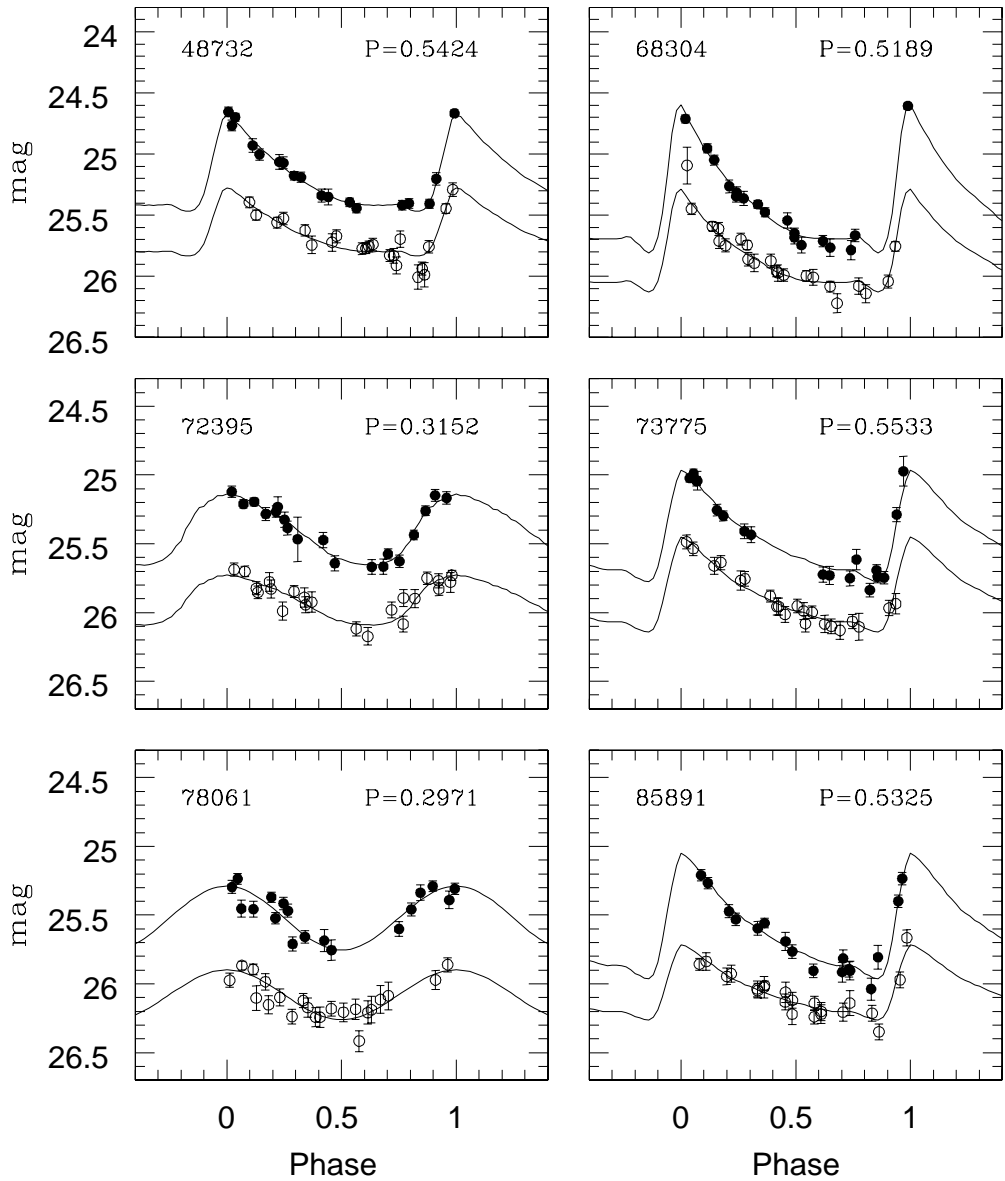,angle=0,clip=,width=18cm}}}
\caption[]{Each panel shows the phased light curve of an RR Lyrae
variable in M33 from the study of Yang et al. (2010). The filled circles
are the F814W points while the open circles are those measured
in F606W. The solid lines are the best fit template to each star.
The periods are indicated in each panel. } 
\label{sarajedini-fig1} 
\end{figure}

\begin{figure}
\centerline{\hbox{\psfig{figure=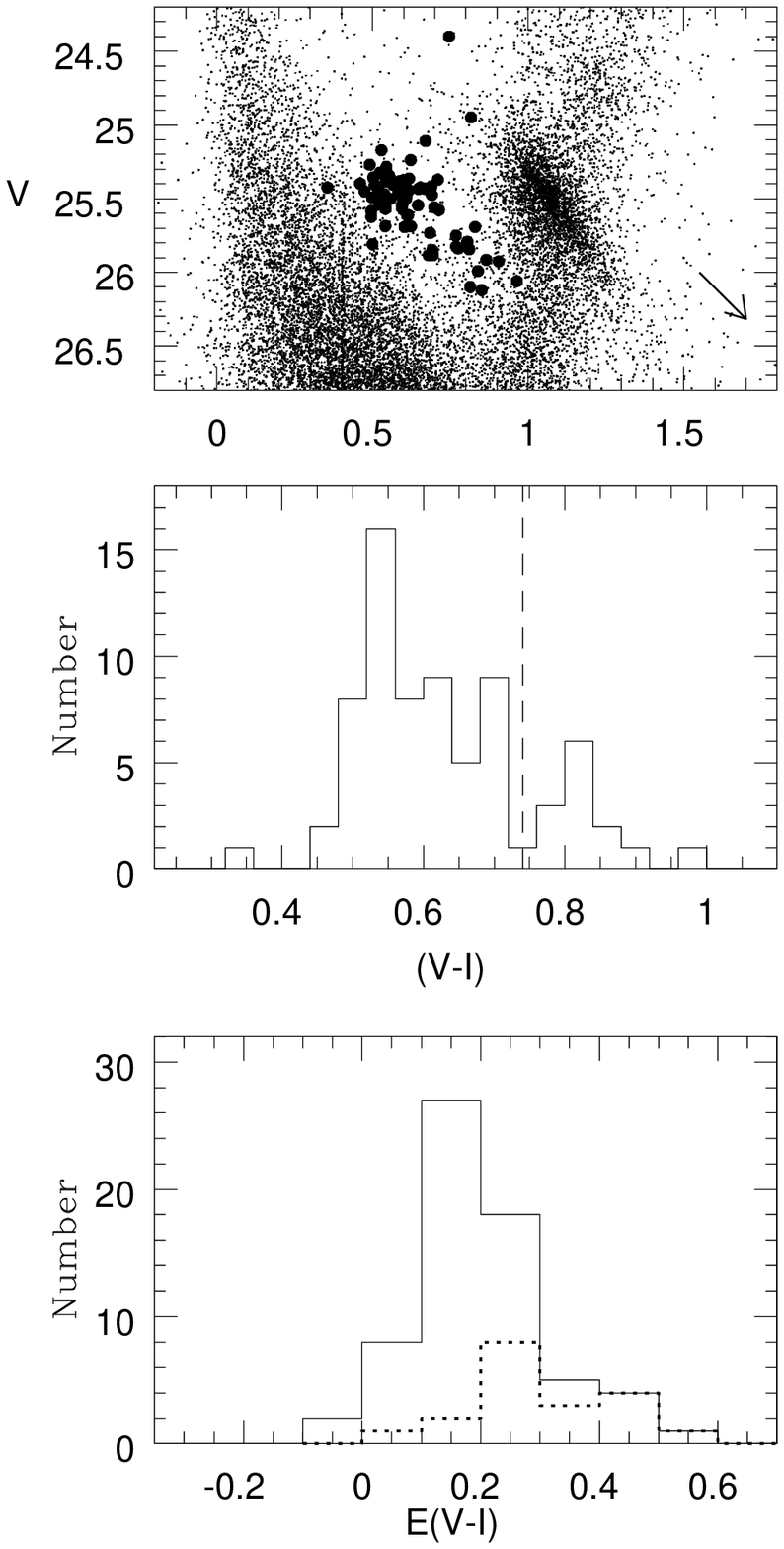,angle=0,clip=,width=16cm}}}
\caption[]{The upper panel shows the color magnitude diagram of
the DISK2 field from the work of Yang et al. (2010) along with the
RR Lyraes identified in this field (filled circles). The brightest two or three stars
are likely to be anomalous cepheids. The arrow shows the reddening
vector in the CMD. The middle panel illustrates the color histogram
for the RR Lyrae variables, which appears to show a bimodal 
distribution with a primary peak at (V--I)$\sim$0.5 and a secondary
one at (V--I)$\sim$0.8. The result of dividing the sample of ab-type 
RR Lyraes at (V--I)=0.74 (dashed line) and plotting the distribution of
reddenings (see text) is shown in the bottom panel. The
thin solid line represents all ab-type RR Lyraes while the 
dotted line shows only those with (V--I)$>$0.74. } 
\label{sarajedini-fig2} 
\end{figure}

\section{Latest Results}

Building upon the work of Sarajedini et al. (2006), Yang et al. (2010)
present an analysis of new HST/ACS/WFC imaging that is part of 
program GO-10190. The primary aim of this program is to study
the star formation of the disk of M33 (Williams et al. 2009). 
As such, fields were obtained
at four different disk locations roughly equally spaced along the 
major axis of M33. Figure 1
of San Roman et al. (2009) shows the locations of the disk fields. Here
we report on the properties of the RR Lyraes in the second closest field
to the center of the M33, which we designate DISK2. 

The observations are composed of 16 epochs in the F606W filter
and 22 in the F814W filter spanning a time window of $\sim$3 days.
Template-fitting analysis suggests the presence of 86 RR Lyraes
in this field - 65 ab-type, 18 c-type, and 3 d-type variables. Figure 1
shows a montage of some of the light curves. The upper panel of
Figure 2 shows the CMD of DISK2 along with the locations of the
RR Lyraes. The middle panel of Fig. 2 displays the distribution in color
of the ab-type RR Lyraes revealing the presence of two peaks - 
a primary peak at (V--I)$\sim$0.5 and a secondary
one at (V--I)$\sim$0.8. We would like to know if this bimodality is
due to reddening internal to M33. As such, we determine the
reddening of each RR Lyrae using Sturch's method as described in Sec. 1
and then examine the distribution of reddenings to see if the fainter/redder
RR Lyraes, those with (V--I)$>$0.74, do indeed suffer from higher reddening.
The lower panel of Fig. 2 illustrates this effect. The
thin solid histogram represents all ab-type RR Lyraes while the
dotted
histogram shows only those with (V--I)$>$0.74 (dashed line in the middle
panel). Our hypothesis seems to be correct - that the fainter/redder 
RR Lyraes are being affected by extinction internal to M33. This
suggests that the variables near (V--I)$\sim$0.5 are on the near
side of M33 while those with (V--I)$\sim$0.8 are in the disk or on the far side.

We now seek to compare the properties of the low reddening and high reddening
RR Lyraes. In particular, how do they compare in the Bailey Diagram?
This is shown in the upper panel of Fig. 3 where we plot the periods
and amplitudes of the DISK2 RR Lyraes (open circles - ab-types,
open triangles - c-types; filled squares - d-types; filled circles - higher
reddening ab-types). These are compared
with the Oosterhoff I and II loci from Clement \& Rowe (2000). We see
that the ab-type RR Lyraes in M33 (both the low reddening and higher
reddening samples)
are consistent with those in Oosterhoff I Galactic globular 
clusters. The lower panel of Fig. 3 shows the period distribution
of the RR Lyraes - the thin solid line represents all of these stars
while the dotted line shows the higher reddening RR Lyraes.
When compared to the period distribution of the low reddening variables,
those with higher reddening exhibit no significant difference in their mean
periods. This suggests that the mean metallicities of these samples
are indistinguishable from each other. This reinforces the
assertion that the low reddening RR Lyraes as well as most of those 
that suffer from higher reddening are likely to be in the halo of M33.


\begin{figure}
\centerline{\hbox{\psfig{figure=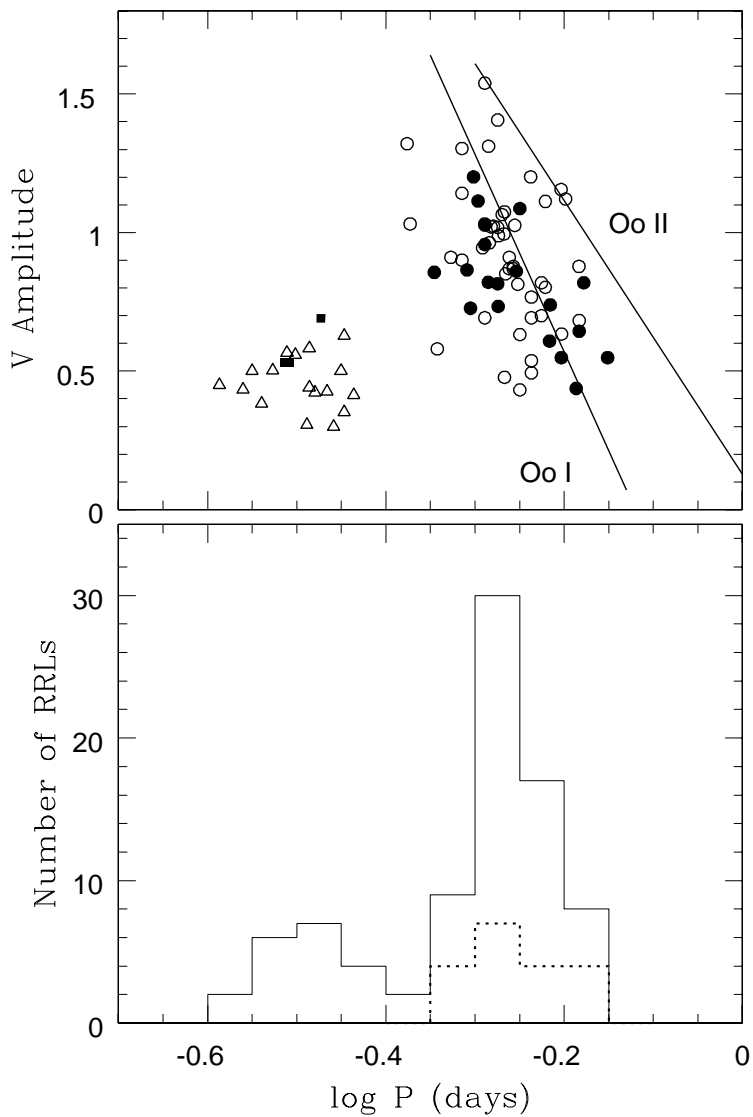,angle=0,clip=,width=16cm}}}
\caption[]{The upper panel is the Bailey Diagram for all RR Lyraes in
the M33 DISK2 field. The open circles are the ab-type variables while
the open triangles are the c-types. The few candidates d-type RR Lyraes
are identified with filled squares. The solid lines are the locations of the
Oosterhoff type I and Oosterhooff type II Galactic globular clusters from
the work of Clement \& Rowe (2000). The filled circles are the ab-type
RR Lyraes with (V--I)$>$0.74 from Fig. 2; these are likely to be located
on the far side of M33's disk. The lower panel shows the period
distribution of the RR Lyraes in the upper panel as the thin line
while the distribution of RR Lyraes with (V--I)$>$0.74 is 
designated by the thick dotted line. See text for a discussion of these
histograms.} 
\label{sarajedini-fig3} 
\end{figure}

\section{Conclusions}

Studies of RR Lyrae variables in the late-type spiral galaxy M33
are still in their infancy but there are a few points we can make
with relative certainty. First, M33 does indeed harbor RR Lyrae
variables in its halo, which suggests that, similar to M31 and the Milky Way,
there is an old component
($\gea$10 Gyr) to the halo of this galaxy. These variables have
pulsational properties that are consistent with those in Oosterhoff I
Galactic globular clusters. Furthermore, their mean metallicity
is [Fe/H]$\sim$-1.4, which is consistent with the mean abundance
of halo globular clusters in M33. Much more work is left to be done.
For example, wide-field time-domain surveys of M33 will reveal the
spatial distribution of the RR Lyraes. The capabilities of 30m telescopes will
likely allow us to obtain kinematic and abundance information for
RR Lyraes in M33. We can then study the detailed properties of the
earliest epochs of star formation in a late type spiral galaxy. This 
has implications for how such galaxies fit into the $\Lambda$CDM
paradigm for structure formation in the early universe. 

\acknowledgements 
The authors gratefully acknowledges the work of a number of close
collaborators that have contributed results to this review.  
The template-fitting code
we use originated with Andy Layden and has been integrated into
our FITLC software by University of Florida
graduate student Conor Mancone. This manuscript has benefited
greatly from a careful reading by Karen Kinemuchi. Much of this work has been
supported by NASA through a
grant from the Space Telescope Science Institute which is operated by 
the Association of
Universities for Research in Astronomy, Incorporated, under NASA 
contract NAS5-26555.


\begin{thebibliography}{}  
\bibitem[alcock2000]{Alcock et al. (2000)} Alcock, C. et al. 2000, \aj, 119, 2194
\bibitem[brown2004]{Brown et al. (2004)} Brown, T. M., Ferguson, H. C.,
Smith, E., Kimble, R. A., Sweigart, A. V., Renzini, A., \& Rich, R. M. 2004,
\aj, 127, 2738
\bibitem[chaboyer1999]{Chaboyer (1999)} Chaboyer, B. 1999, in Post-Hipparcos 
Cosmic Candles, ASSL, Vol. 237, edited by A. Heck and F. Caputo 
(Dordrecht: Kluwer Academic Publishers) p.111
\bibitem[clement2000]{Clement and Rowe (2000)} Clement, C. M.
\& Rowe, J. 2000, \aj, 120, 2579
\bibitem[dotter2009]{Dotter et al. (2009)} Dotter, A. et al. 2009, \apj, in press
(http://arxiv.org/abs/0911.2469)
\bibitem[galleti2004]{Galleti et al. (2004)} Galleti, S., Bellazzini, M.,
\& Ferraro, F. R. 2004, \aap, 423, 925
\bibitem[gratton2003]{Gratton et al. (2003)} Gratton, R. G., Bragaglia, A.,
Carretta, E., Clementini, G., Desidera, S., Grundahl, F., \&
Lucatello, S., \aap, 408, 529
\bibitem[gratton2004]{Gratton et al. (2004)} Gratton, R. G., Bragaglia, A.,
Clementini, G., Carretta, E., Di Fabrizio, L., Maio, M., \& Taribello, E.
\aap, 421, 937
\bibitem[gulden2005]{Guldenschuh et al. (2005)} Guldenschuh, K. et al. 2005, 
\pasp, 117, 721
\bibitem[kunder2009]{Kunder et al. (2009)} Kunder, A., Chaboyer, B.,
\& Layden, A. C. 2009, \aj, in press (http://xxx.lanl.gov/abs/0911.1770)
\bibitem[layden2000]{Layden and Sarajedini (2000)} Layden A. C., \&
Sarajedini, A. 2000, \aj, 119, 1760
\bibitem[mancone2008]{Mancone and Sarajedini (2008)} Mancone, C. L.
\& Sarajedini, A. 2008, \aj, 136, 1913
\bibitem[navarro1997]{Navarro, Frenk, and White (1997)} Navarro, J. F., 
Frenk, C. S., \& White, S. D. M. 1997,  \apj, 490, 493
\bibitem[pritchet1988]{Pritchet (1988)} Pritchet, C. J. 1988, in The 
Extragalactic Distance Scale, ASP Conf. Ser. (ASP: San Francisco) p. 59
\bibitem[sanroman2009]{San Roman et al. (2009)} San Roman, I.,
Sarajedini, A., Garnett, D. R., \& Holtzman, J. A. 2009, \apj, 699, 839
\bibitem[sarajedini2000]{Sarajedini et al. (2000)} Sarajedini, A., 
Geisler, D., Schommer, R. \& Harding, P. 2000, \aj, 120, 2437
\bibitem[sarajedini2006]{Sarajedini et al. (2006)} Sarajedini, A., 
Barker, M. K., Geisler, D., Harding, P., \& Schommer, R. 2006, \aj, 132, 1361
\bibitem[sarajedini2009]{Sarajedini et al. (2009)} Sarajedini, A.,
Mancone, C., Lauer, T. R., Dressler, A., Freedman, W., Trager, S. C.
Grillmair, C., \& Mighell, K. J. 2009, \aj, 138, 184
\bibitem[smith1995]{Smith (1995)} Smith, H. in RR Lyrae Stars, Cambridge
Astrophysics Series, (Cambridge University Press: Cambridge) p. 14
\bibitem[sturch1966]{Sturch (1966)} Sturch, C. \apj, 143, 774
\bibitem[williams2009]{Williams et al. (2009)} Williams, B. F., 
Dalcanton, J. J., Dolphin, A. E., Holtzman, J., \& Sarajedini, A. 2009,
\apj, 695, L15
\bibitem[yang2010]{Yang et al. (2010)} Yang, S. -C. et al. 2010, in
preparation.
\end{thebibliography}
\end{document}